\PassOptionsToPackage{table,dvipsnames}{xcolor}
\documentclass[sigconf]{acmart}

\AtBeginDocument{%
  \providecommand\BibTeX{{%
    \normalfont B\kern-0.5em{\scshape i\kern-0.25em b}\kern-0.8em\TeX}}}

\definecolor{customgray}{HTML}{EAE5EB}
\definecolor{navy}{RGB}{0,64,158}

\usepackage{xcolor}
\usepackage{amsmath}
\usepackage{graphicx}
\usepackage{booktabs}
\usepackage{array}
\graphicspath{{Figure/}} 
\usepackage{enumitem}

\copyrightyear{2026}
\acmYear{2026}
\setcopyright{cc}
\setcctype{by-nc-nd}
\acmConference[CHI EA '26]{Extended Abstracts of the 2026 CHI Conference on Human Factors in Computing Systems}{April 13--17, 2026}{Barcelona, Spain}
\acmBooktitle{Extended Abstracts of the 2026 CHI Conference on Human Factors in Computing Systems (CHI EA '26), April 13--17, 2026, Barcelona, Spain}
\acmDOI{10.1145/3772363.3799044}
\acmISBN{979-8-4007-2281-3/2026/04}

\begin{document}

\title{Modeling Sequential Design Actions as Designer Externalization on an Infinite Canvas
}

\author{Yejin Yun}
\authornote{Co-first authors}
\affiliation{
  \department{Design Informatics Lab, Interior Architecture Design}
  \institution{Hanyang University}
  \city{Seoul}
  \country{Republic of Korea}
}
\affiliation{
  \department{Human-Centered AI Design Institute}
  \institution{Hanyang University}
  \city{Seoul}
  \country{Republic of Korea}
}
\email{carryy0823@gmail.com}

\author{Seung Won Lee}
\authornotemark[1]
\affiliation{
  \department{Design Informatics Lab, Interior Architecture Design}
  \institution{Hanyang University}
  \city{Seoul}
  \country{Republic of Korea}
}
\affiliation{
  \department{Human-Centered AI Design Institute}
  \institution{Hanyang University}
  \city{Seoul}
  \country{Republic of Korea}
}
\email{lswgood0901@gmail.com}

\author{Jiin Choi}
\authornotemark[1]
\affiliation{
  \department{Design Informatics Lab, Interior Architecture Design}
  \institution{Hanyang University}
  \city{Seoul}
  \country{Republic of Korea}
}
\affiliation{
  \department{Human-Centered AI Design Institute}
  \institution{Hanyang University}
  \city{Seoul}
  \country{Republic of Korea}
}
\email{jiin4900@gmail.com}

\author{Kyung Hoon Hyun}
\authornote{Corresponding author.}
\affiliation{
  \department{Design Informatics Lab, Interior Architecture Design}
  \institution{Hanyang University}
  \city{Seoul}
  \country{Republic of Korea}
}
\affiliation{
  \department{Human-Centered AI Design Institute}
  \institution{Hanyang University}
  \city{Seoul}
  \country{Republic of Korea}
}
\email{hoonhello@gmail.com}

\begin{abstract}
Infinite canvas platforms are becoming central to contemporary design practice, enabling designers to externalize cognition through the spatial arrangement of multimodal artifacts. As AI agents increasingly generate and organize content within these environments, their impact on designers' externalization processes remains underexplored. We report a field study with eight professional designers comparing workflows with and without an AI organizing agent. Through a sequence analysis of 5,838 design actions, we identify three key shifts: (1) AI integration reallocates cognitive effort from spatial management to content curation and relational structuring, without increasing active time; (2) a characteristic generate-and-curate cycle emerges in which designers’ demands on the agent intensify while the agent’s functional role adapts; and  (3) AI’s role evolves from a divergent catalyst in early stages to a convergent curator in later phases. These findings offer a behavioral model for designing phase-adaptive AI tools that support human–AI co-evolution on infinite canvases.
\end{abstract}



\maketitle
\begin{figure*}[t]
\centering
\includegraphics[width=\linewidth]{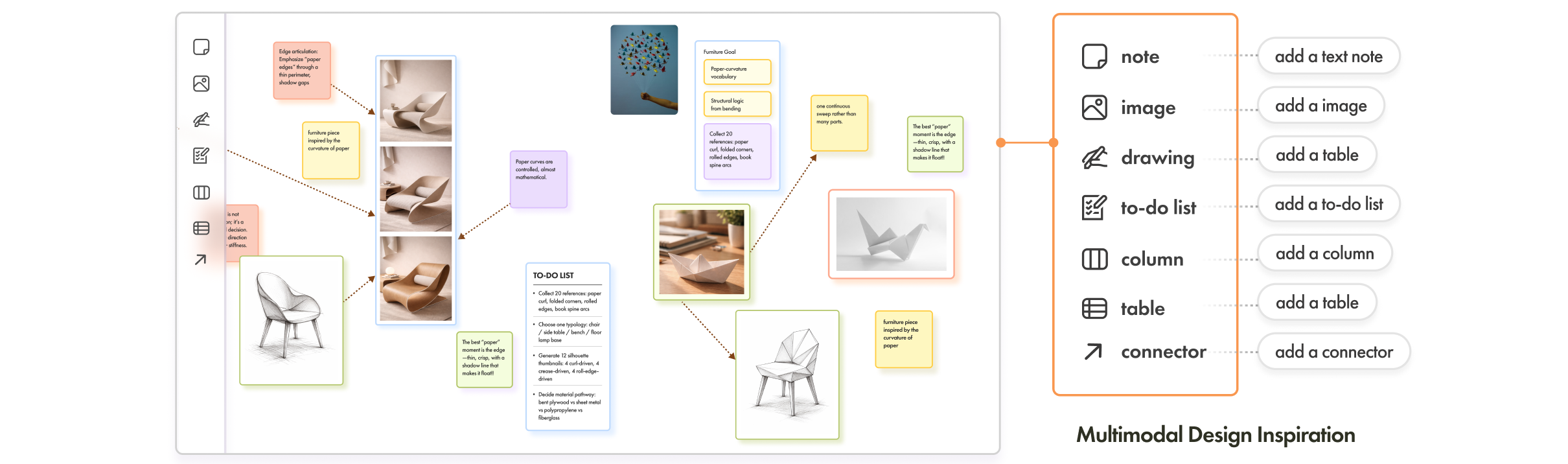}
\caption{Infinite-canvas workspace used in the study. Designers externalize ideas by collecting, spatially arranging, and connecting multimodal artifacts (e.g., images, notes, connectors). Both Baseline and Agent$_{\text{organizer}}$ conditions support prompt generation and image editing within the canvas.}
\Description{Illustration of the infinite-canvas design workspace used in the study. Multimodal artifacts—images, sticky notes, and text blocks—are spatially arranged and connected with lines across the canvas. A panel on the right lists artifact types (note, image, drawing, to-do list, column, table, connector), illustrating how designers externalize ideas by collecting, organizing, and linking visual and textual materials.}
\label{fig:interface}
\end{figure*}

\section{Introduction}
Contemporary design practice is progressively migrating toward infinite canvas platforms. An infinite canvas is a boundless digital workspace where designers externalize their thinking by collecting multimodal artifacts—notes, images, and drawings—and constructing semantic connections through spatial arrangement\cite{Croisdale_2025, guo2026protosampling}. Unlike linear tools, these environments capture design reasoning through the spatial configuration of artifacts, producing “cognitive traces” that reveal how designers navigate from uncertainty toward solutions\cite{Bentvelzen_2022}. The inspirational materials, structured compositions, and connectors left on the infinite canvas are not mere byproducts but continuous records of sensemaking\cite{Zhu_2024, gorkovenko2023data, Anderson_2025}, enabling designers to revisit, revise, reconfigure, and reflect upon evolving ideas.
Despite the potential of infinite canvases for observing design cognition, prior work has mainly evaluated usability or final outcomes, leaving process-level accounts relatively limited\cite{Croisdale_2025, cox2025beyond}. In AI-assisted design, systems have begun to support the collection and organization of sticky notes (or similar externalizations\cite{chen2026stickynexus, Ball_2021}), yet log-based evidence on how such support shapes in-canvas behavior, such as action sequences or loops and phase transitions from ideation to organization and elaboration, remains scarce. This gap is widening with AI organizing agents that automatically cluster artifacts\cite{chen2026stickynexus, Yun_2025} and generate contextual content\cite{guo2026protosampling, Croisdale_2025}, but how these agents alter temporal rhythms, behavioral sequences, and phase-specific strategies is still not well understood.

To address this gap, we conducted a field study with eight professional designers, comparing workflow patterns across two conditions with and without an AI organizing agent. In both conditions, designers developed ideas by collecting and accumulating multimodal artifacts on the infinite canvas, spatially arranging them, and establishing connections. To systematically identify how the AI organizing agent affects designers’ externalization processes, we formulated three research questions:

\begin{itemize}
    \item RQ1. How does the AI organizing agent reconfigure the temporal structure and action distribution of design workflows?
    \item RQ2. What new sequential design loops emerge during AI collaboration, and how do they replace existing spatial management practices? 
    \item RQ3. How does the AI organizing agent’s role evolve across design process stages (early, mid, and late)?
\end{itemize}
To address these questions, we developed a \textbf{Sequential Design Action Taxonomy} that maps raw interaction logs to semantically meaningful design action types. This taxonomy provides the analytical foundation for systematically examining workflow dynamics and answering each research question.

We summarize our contributions as follows. First, \textbf{we propose an externalization-centered behavioral model for infinite canvases.} We conceptualize designers' externalization as observable action primitives over multimodal artifacts, enabling computational models of how thinking becomes spatial representation. Second, \textbf{we introduce a sequence-based analysis pipeline for externalization dynamics.} We build a pipeline that abstracts raw interaction logs into semantic action sequences, enabling quantitative comparisons across conditions and identifying discriminative sequential patterns. Third, \textbf{we present empirical findings of phase-dependent human–AI collaboration.} We demonstrate that the proactive organizing agent redistributes cognitive effort from spatial management to content curation and relational structuring without increasing workload, and that its functional role evolves across design phases from idea amplifier to co-evolutionary structuring partner to convergent curator, following a non-monotonic pattern in which designers' generation selectivity and acceptance commitment increase simultaneously.

\begin{figure*}[t]
\centering
\includegraphics[width=\linewidth]{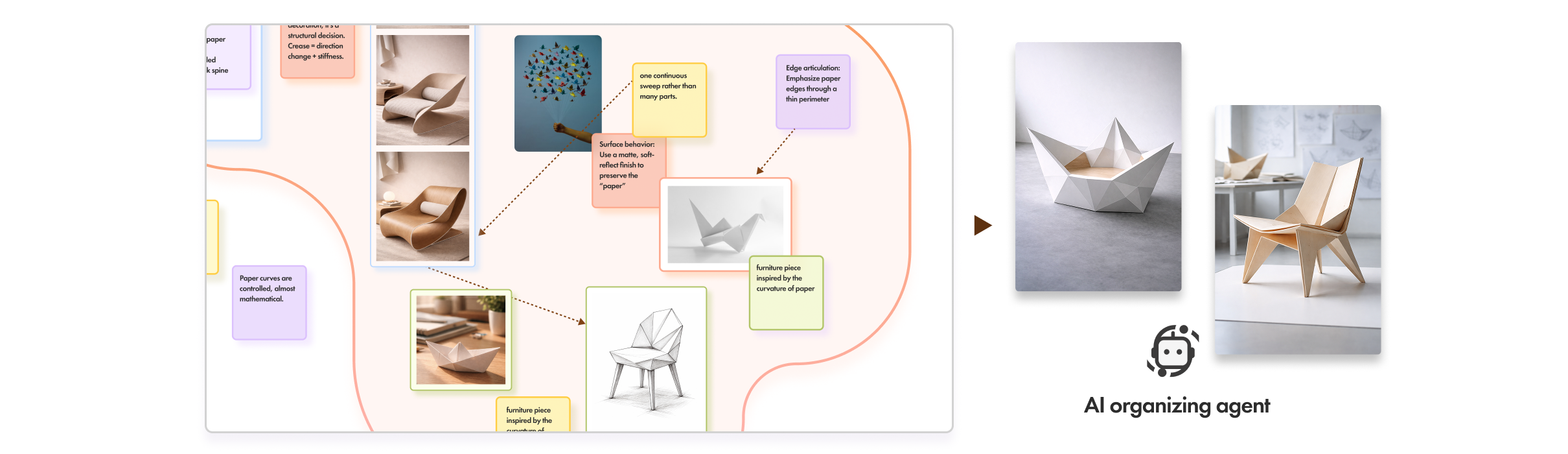}
\caption{Workflow of the Agent$_{\text{organizer}}$ condition. The AI organizing agent observes designers' artifact collections on the canvas, forms semantically coherent clusters, and leverages these clusters as implicit prompts for subsequent image generation.}
\Description{Illustration of the Agent_organizer condition. The canvas contains clustered multimodal artifacts (images and notes) connected by lines, indicating how the AI agent observes designers’ artifact collections and organizes them into semantic groups. Generated images appear on the right, showing how these clusters are used as implicit prompts for subsequent image generation.}
\label{fig:agent}
\end{figure*}

\section{Study Design}
We conducted a field study with eight professional designers (product and space design; mean years of experience = 4.5, SD = 2.20) to examine how interaction patterns and design actions change with Agent$_{\text{organizer}}$. Participants alternated between Baseline (infinite canvas for manually collecting/arranging multimodal artifacts such as text and images, with prompt generation and image editing) (Figure \ref{fig:interface}) and Agent$_{\text{organizer}}$ (same canvas and AI features plus an organizing agent that clusters existing artifacts and uses curated groups as visual references for generating new images)(Figure \ref{fig:agent}). Each designer completed two tasks per condition: a Space brief (Tea lounge: compact evening retreat with subdued tone, tactile comfort, and soft safety-minded details) and a Product brief (Kitchenware: everyday tools for minimal households emphasizing ergonomic clarity, durable tactility, and a restrained, cohesive material expression), with task–condition order counterbalanced across participants. Timestamped interaction logs were collected for analysis; event types and preprocessing are described in the Supplementary Material.

\begin{table*}[t]
\centering
\caption{Hierarchical design action taxonomy with 11 broad categories, seven artifact types, and representative action examples. Categories are derived through state-based inference (Create–Interact) or explicit system events (AgentGen–IntentEdit).}
\label{tab:action_taxonomy}
\small
\setlength{\tabcolsep}{0.5pt}
\renewcommand{\arraystretch}{0.8}

\begin{tabular}{@{}%
  >{\raggedright\arraybackslash}p{2cm}%
  >{\raggedright\arraybackslash}p{4cm}%
  >{\raggedright\arraybackslash}p{2.8cm}%
  >{\raggedright\arraybackslash}p{7.2cm}%
@{}}
\toprule
\textbf{Category} & \textbf{Meaning} & \textbf{Artifact type} & \textbf{Example actions} \\
\midrule
Create & Creating a new artifact & All & \texttt{Create\_inspiration}, \texttt{Create\_inspiration\_agent}, \texttt{Create\_structure\_column} \\
\midrule
Elaborate & Refining content & All & \texttt{Elaborate\_inspiration\_micro}  \\
\midrule
Relocate & Spatial repositioning & All & \texttt{Relocate\_inspiration\_micro}, \texttt{Relocate\_structure\_large} \\
\midrule
Relate & Connector management & Relation & \texttt{Relate} \\
\midrule
Structure & Hierarchy organization & Container & \texttt{Structure} \\
\midrule
Prune & Deleting artifacts & All & \texttt{Prune\_inspiration}, \texttt{Prune\_structure} \\
\midrule
Interact & Social engagement & Comments, reactions & \texttt{Interact\_comment\_create}, \texttt{Interact\_reaction} \\
\midrule
AgentGen & AI agent organizing/generation & Agent & \texttt{Agent\_gen} \\
\midrule
PromptGen & User prompt generation & AI mode & \texttt{Prompt\_gen} \\
\midrule
ImageEdit & AI in editing & Image & \texttt{Image\_edit} \\
\midrule
IntentEdit & Aligning design intent & Intent & \texttt{Intent\_edit\_global}, \texttt{Intent\_edit\_local} \\
\bottomrule
\end{tabular}
\end{table*}

\section{Sequential Design Action Classification Framework}
\label{section:3}
After preprocessing 14,075 raw interaction events with a rule-based Event Log Abstraction (ELA) pipeline—removing operational noise and aggregating micro-events into 5,838 analyzable \textit{Design Actions}, we developed a computational framework that classifies these actions into a hierarchical taxonomy. The taxonomy distinguishes state-based inference from explicit system events and comprises 11 broad categories (\textit{Create, Elaborate, Relocate, Relate, Structure, Prune, Interact, AgentGen, PromptGen, ImageEdit, IntentEdit}), seven artifact types \textit{(note, image, drawing, table, column, to-do list, and connector)}, and 37 or more specific action types. The classification pipeline (Figure \ref{fig:interface}) processes each timestamped event through three sequential stages.

\noindent \textbf{Stage 1: Change detection.} In this stage, the system detects technical-level state changes by comparing each artifact instance's current state (S$_{\text{t}}$) with its previous state (S$_{\text{t-1}}$). Specifically, six types of deltas are extracted: (1) instance appearance or removal, (2) character-level changes in text content ($\Delta_{\text{char}}$), (3) size changes in visual elements ($\Delta_{\text{size}}$: width/height), (4) position changes measured by Euclidean distance ($\Delta_{\text{pos}}$), (5) increases in social signals (reaction counts, comments), (6) modifications in relationship endpoints and hierarchical container (startId/endId, child ID changes). These low-level primitives purely quantify state differences without semantic interpretation. Full delta definitions and details are provided in the Supplementary Material, Section A.2.

\noindent \textbf{Stage 2: Hierarchical Taxonomy mapping.} We employed a hybrid classification strategy to derive 11 broad action categories. Seven categories (\textit{Create, Elaborate, Relocate, Relate, Structure, Prune, Interact}) are inferred by analyzing geometric and semantic deltas between artifact states. Four AI-mediated categories (\textit{AgentGen, PromptGen, ImageEdit, IntentEdit}) are identified directly from system task codes and metadata logs.

To ensure consistent categorization across heterogeneous objects, we standardized artifacts into four groups: Inspiration (images, notes, drawings, agent-generated images), Container (columns, tables), Control (to-do lists), and Relation (connectors).

Each detected event is mapped to one broad category and a specific action subtype. \textit{Create} captures new artifact instances and functions as an anchoring operation; \textit{Elaborate} captures content refinement through character-level text edits; \textit{Relocate} captures spatial repositioning discretized by magnitude; \textit{Relate} manages connectors including creation, reconnection, and attribute changes; \textit{Structure} organizes hierarchical containers; \textit{Prune} removes artifacts or structures; and \textit{Interact} tracks artifact level interaction (e.g., reactions and comments). AI-mediated categories are derived from explicit system logs, with \textit{AgentGen} occurring only in the Agent$_{\text{organizer}}$ condition.

\begin{figure*}[t]
\centering
\includegraphics[width=\linewidth]{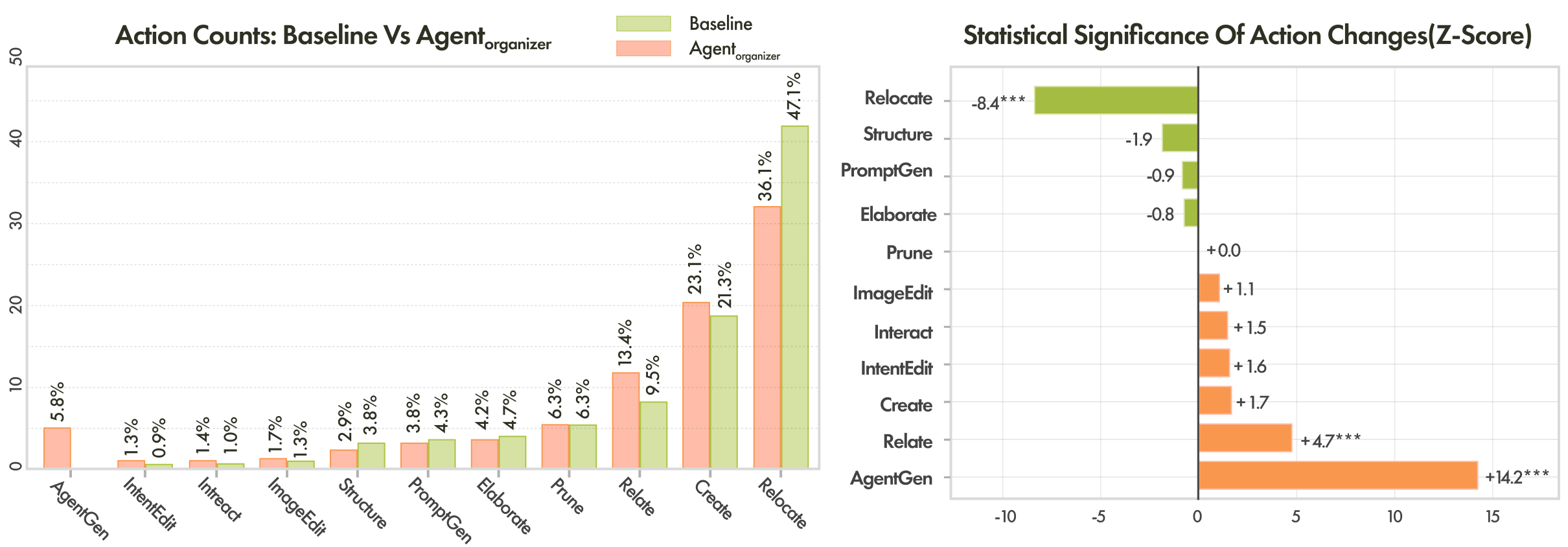}
\caption{Comparative analysis of design action distributions and statistical significance between Baseline and Agent$_{\text{organizer}}$ conditions. 
\textbf{Left:} Percentage distribution of 11 design actions across conditions. \textbf{Right:} Z-scores indicating statistically significant changes in action frequency, highlighting decreases in Relocate and increases in AgentGen and Relate.}
\Description{Two-panel chart comparing design action distributions between the Baseline and Agent_organizer conditions. The left panel shows bar charts of percentage distributions for each action category, indicating that Relocate decreases while AgentGen appears only in the agent condition. The right panel shows Z-scores of action frequency changes, highlighting significant decreases in Relocate and significant increases in AgentGen and Relate.}
\label{fig:action_comparison}
\end{figure*}

\section{Results and Discussion
}
\subsection{Design Workflow Redistribution Without Increased Effort (RQ1)}
Total active time did not differ between conditions (Baseline: $M = 3.73$ h, $SD = 1.75$; Agent$_{\text{organizer}}$: $M = 4.03$ h, $SD = 2.13$; Wilcoxon signed rank $W = 16.0$, $p = .84$). However, action distribution changed substantially ($\chi^2(10, N = 5{,}838) = 269.59$, $p < .001$, Cramer's $V = 0.215$). The most pronounced shift occurred in \textit{Relocate}, which dropped from $47.11\%$ to $36.11\%$, a reduction of $11.0$ percentage points that corresponded to the increase in \textit{AgentGen} ($5.84\%$). Sensitivity analysis excluding \textit{AgentGen} confirmed that the redistribution remained significant ($\chi^2(9, N = 5{,}693) = 69.50$, $p < .001$), indicating that the agent did not simply add new actions but restructured how designers allocated their effort (Figure \ref{fig:action_comparison}). Notably, a significant redistribution target was \textit{Relate}, which rose from $9.45\%$ (Baseline) to $13.41\%$ (Agent$_{\text{organizer}}$; $Z = +4.75$, $p < .001$), indicating that designers in the Agent$_{\text{organizer}}$ condition spent proportionally more effort establishing semantic relationships between artifacts.

This pattern constant engagement with redistributed activity aligns with Shneiderman’s Human-Centered AI framework, which distinguishes augmentation that amplifies human capabilities from automation that replaces human effort\cite{shneiderman2020human}. The agent did not reduce designers' cognitive investment; it redirected that investment toward different types of design work.  Interview accounts corroborated this pattern: designers described baseline workflows as requiring substantial manual effort for basic grouping (P4: ``planning from a blank slate''), whereas agent-mediated clustering absorbed much of this overhead (P6), freeing capacity that was redirected toward relational structuring. This result suggests that proactive AI intervention reshape what designers do rather than how much they do—a form of strategic transformation that the following sections discuss in detail. 

\begin{figure*}[t]
\centering
\includegraphics[width=\linewidth]{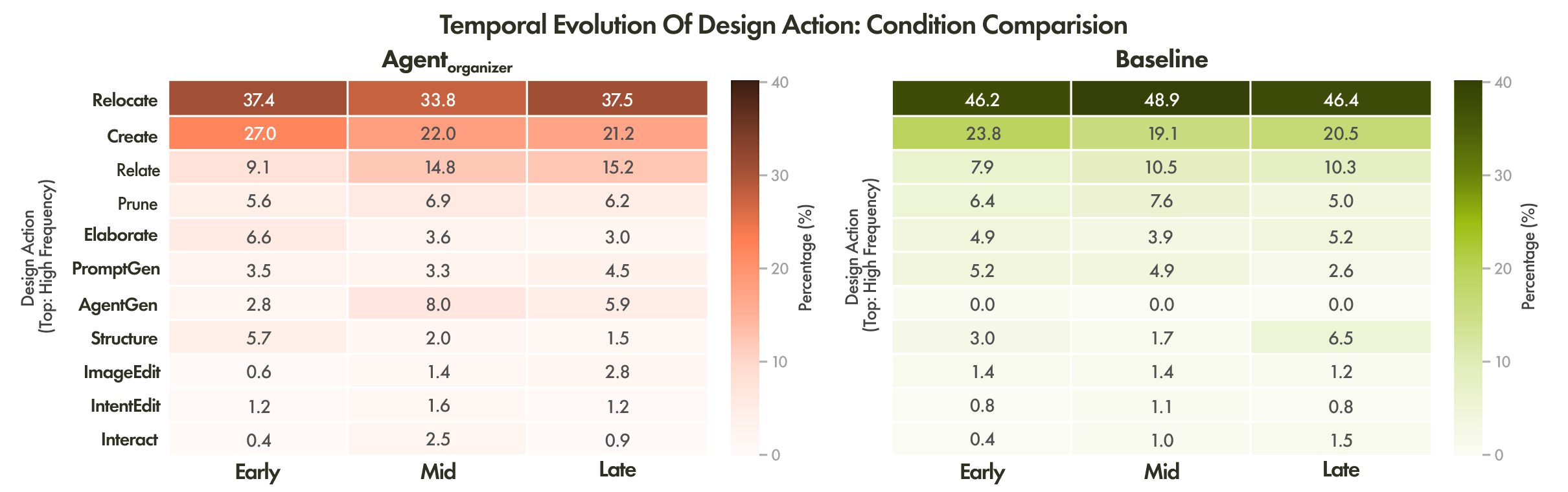}
\caption{Temporal evolution of design actions across Early, Mid, and Late phases Agent$_{\text{organizer}}$ vs. Baseline. Heatmaps show the percentage distribution of top design actions over time, revealing reduced Relocate activity and increased content creation and relational structuring in the agent condition.}
\Description{Two heatmaps comparing the temporal distribution of design actions across Early, Mid, and Late phases in the Agent_organizer and Baseline conditions. Each row represents a design action and each column a phase, with color intensity indicating percentage frequency. The visualization shows that Relocate remains the dominant action across phases but is consistently lower in the agent condition, while actions such as Create and Relate increase in the Agent_organizer condition, indicating a shift toward content creation and relational structuring over time.}
\label{fig:sequence}
\end{figure*}

\subsection{Sequential design action mining, From Spatial Management Overhead to Creative Flow: Shift in Collaborative Loops (RQ2)}
In the Baseline condition, “Spatial management overhead” dominated the workflow: The \textit{Relocate} $\rightarrow$ \textit{Relocate} self-loop accounted for $29.05\%$ of all Baseline bigrams, significantly exceeding the Agent$_{\text{organizer}}$ condition ($18.98\%$, $Z = -8.80$, $p < .001$). This dominance extended to the trigram level, where \textit{Relocate} appeared in six of the eight trigram patterns before a non-\textit{Relocate} chain, \textit{Relate} $\rightarrow$ \textit{Relate} $\rightarrow$ \textit{Relate}, emerged at Rank~9 ($2.04\%$). In the Agent$_{\text{organizer}}$ condition, designers were freed from this overhead: \textit{Relate} $\rightarrow$ \textit{Relate} $\rightarrow$ \textit{Relate} rose to Rank~3 ($2.88\%$), immediately following the two self-repetition patterns common to both conditions (\textit{Relocate}$\times3$ and \textit{Create}$\times3$). The \textit{Relocate}$\times3$ trigram itself declined from $20.85\%$ to $12.76\%$, a relative reduction of $38.80\%$, indicating that the agent created sequential room for sustained relational work that spatial coordination had previously occupied. This freed capacity was partly absorbed by new AI-mediated loops: \textit{AgentGen} $\rightarrow$ \textit{Create} and \textit{AgentGen} $\rightarrow$ \textit{AgentGen} entered the top 10 bigrams (Rank~9 and 10, respectively), patterns entirely absent in the Baseline condition. 

Designers confirmed this shift qualitatively: Several reported that agent support changed their entry point from manual keyword derivation to reference-driven exploration (P1, P2), and others strategically adjusted their recording practices to influence downstream agent behavior. For instance, becoming more deliberate about which records to upload (``thinking about what the agent would see,'' P5).

Zhou et al. identified a tension between linear command-execution AI workflows and the nonlinear nature of creative design \cite{zhou2024understanding}. Both conditions in our study included AI generation tools, but only Agent$_{\text{organizer}}$ introduced proactive intervention, that is, clustering artifacts and generating contextual content without explicit user commands. Our findings suggest that this proactive engagement absorbed spatial management overhead, creating cognitive room for designers to evaluate content and align design intent rather than coordinate artifact placement.

\subsection{Phase-Dependent Evolution of Agent Partnership (RQ3)}
Sequential design behavior patterns within each phase, analyzed through first-order Markov chains, revealed a distinct evolution in the AI organizing agent's roles. The reduction in spatial management overhead described in Section~4.2 was not evenly distributed across phases of the design timeline. $P(\textit{Relocate}|\textit{Relocate})$ in the Baseline remained consistently high across all phases ($61.51\% \rightarrow 68.85\% \rightarrow 58.89\%$), whereas the Agent$_{\text{organizer}}$ condition showed substantially lower values ($49.89\% \rightarrow 52.55\% \rightarrow 58.44\%$), with the gap narrowing progressively from an 11.62 percentage-point difference in the Early phase to near convergence in the Late phase. This narrowing gap suggests that the agent’s contribution shifted over time; the capacity it freed was not always directed toward the same activities. Phase-level transition and action-rank analyses revealed three distinct roles the agent assumed as the design process unfolded.

\paragraph{Early phase, Idea Amplifier.}
The Early phase was characterized by the agent functioning as a catalyst for divergent exploration. This aligns with research on divergent ideation through parallel prototyping~\cite{dow2009efficacy}. The self-transition probability $P(\textit{AgentGen}|\textit{AgentGen})$ was $33.33\%$, the lowest across phases, whereas $P(\textit{Create}|\textit{AgentGen})$ (i.e., \textit{AgentGen} $\rightarrow$ \textit{Create}) reached $55.56\%$, the highest. Rather than sustaining iterative generation, the agent’s outputs rapidly triggered new artifact placement. Detailed action-level analysis revealed that $60.00\%$ of placements were agent-generated images accepted directly, whereas $40.00\%$ involved designers sourcing their own artifacts after bypassing the agent's suggestion, the highest rate of independent sourcing across all phases. This influx of new material coincided with shifts in activity patterns: \textit{Relate} increased from $7.87\%$ in the Baseline condition to $9.10\%$ in Agent$_{\text{organizer}}$, and \textit{Elaborate} increased from $4.92\%$ to $6.61\%$, rising from Rank~6 to Rank~4. Notably, \textit{Prune}, which was ranked fourth in the Baseline condition, dropped to Rank~6 in the Agent$_{\text{organizer}}$ condition, indicating that \textit{Elaborate} replaced \textit{Prune} at this position. \textit{Elaborate} reached its highest proportion in this phase, suggesting that designers used the initial wave of agent-generated and self-sourced raw material for content refinement. The agent’s proposals enabled designers to populate the canvas with diverse artifacts and begin refining individual elements. Consistent with this catalytic role, one designer contrasted agent-generated ``unexpected, refreshing surprises'' with baseline outcomes that ``came out exactly as initially intended'' (P2), and another reported that agent support provided directional scaffolding even in an unfamiliar design domain (P1).

\paragraph{Mid phase, Co-evolutionary Structuring.} In the Mid phase, the self-transition probability $P(\textit{AgentGen}|\textit{AgentGen})$ increased sharply to $41.89\%$ (Rank~1), indicating that designers engaged in more sustained generation cycles, requesting multiple alternatives before selecting. However, $P(\textit{Create}|\textit{AgentGen})$ dropped to $37.84\%$, the lowest value across phases, revealing that designers became more selective about which outputs reached the canvas. Despite this lower placement rate, the acceptance ratio (i.e., the proportion of agent-generated outputs accepted and placed as Create\_inspiration\_agent) rose to $89.29\%$, demonstrating a transition from designer-initiated to agent-mediated artifact addition—an alignment suggesting that the agent's solutions have synchronized with the designers' evolving problem definition~\cite{dorst2001creativity}. This selectivity was accompanied by a marked increase in relational organization: \textit{Relate} as a standalone action rose from $9.10\%$ (Early) to $14.83\%$, and the \textit{Relate} $\rightarrow$ \textit{Relate} bigram doubled from $3.33\%$ (Rank~7) to $6.39\%$ (Rank~4), indicating sustained relational chaining. These patterns suggest that Mid-phase designers shifted their cognitive focus from idea generation to emerging organizational structure, demanding more alternatives from the agent while selectively integrating only those that fit an emerging organizational structure. Accordingly, the agent’s role shifted from an inspiration catalyst to relational structuring partner~\cite{cross2023design}.

\paragraph{Late phase, Convergent Curation.} The Late phase was defined by convergent integration and curation. The self-transition probability $P(\textit{AgentGen}|\textit{AgentGen})$ reached its peak at $43.14\%$ (Rank~2), reflecting the most intensive iterative generation. $P(\textit{Create}|\textit{AgentGen})$ rebounded to $45.10\%$ (Rank~1), completing a U-shaped trajectory ($55.56\% \rightarrow 37.84\% \rightarrow 45.10\%$), and the acceptance rate climbed to $95.65\%$, indicating near-complete reliance on agent-generated artifacts for canvas additions. The proportion of \textit{Create} as a standalone action reached its lowest level across phases ($27.02\% \rightarrow 21.99\% \rightarrow 21.25\%$), further confirming the shift from designer-initiated to agent-mediated content creation. At the same time, \textit{Relate} peaked at $15.24\%$, and the \textit{Relate} $\rightarrow$ \textit{Relate} bigram reached $7.20\%$ (Rank~2, the highest across phases), signaling intensive consolidation of artifact relationships. Together, these patterns suggest that the Late phase was characterized by convergent curation: Designers leveraged the agent for targeted content generation with near-universal acceptance, while simultaneously intensifying relational connections to stabilize their design vision. Designers' reflections echoed this convergent pattern: P5 evaluated agent-condition outcomes as simultaneously more aligned with client requirements and more creatively distinctive than baseline results, despite initially perceiving agent suggestions as overly directive.

\section{Conclusion and Limitation
}
This paper presents a sequence-based behavioral model of design externalization on infinite canvases. Through analysis of 5{,}838 design actions, we demonstrated three key findings. Agent$_{\text{organizer}}$ reallocated cognitive effort from spatial management to content curation and relational structuring without increasing active time, indicating strategic redistribution rather than added workload. Second, the agent’s role evolved across a three-phase trajectory: as an inspirational catalyst whose proposals stimulated canvas population and also increased \textit{Elaborate} and \textit{Relate} activity (Early), a co-evolutionary structuring partner where designers demanded more alternatives yet curated selectively while doubling relational work (Mid), and a convergent curator where near-universal acceptance and peak relational chaining consolidated the design vision (Late). The U-shaped $P(\textit{Create}|\textit{AgentGen})$ pattern and the progressive acceptance shift reveal that human--AI collaboration on infinite canvases is non-monotonic: Designers develop increasing selectivity in their requests to the agent, while increasing trust in what they ultimately accept, a behavioral tension resolved by the agent’s shifting role from idea amplifier to selective curator. 

These findings suggest that the value of generative AI in infinite canvas environments lies not in automation but in phase-adaptive, bidirectional collaboration. This phase-dependent evolution demonstrates that AI agents function as dynamic partners adapting to designers’ stage-specific needs rather than uniform tools, where designers’ multimodal archiving behavior functions as implicit prompts, reducing reliance on explicit commands. Accordingly, future AI canvas systems should align their affordances with this evolving cognitive rhythm: supporting inspirational breadth during early exploration, selective scaffolding during mid-phase structuring, and convergent integration during late-phase curation. In this sense, proactive AI should be understood not as a substitute for human agency, but as a mechanism for redistributing cognitive effort across phases of design. Although these patterns were derived from a study with eight designers, limiting generalizability, they offer a foundation for validating phase-dependent human–AI collaboration models across larger samples and broader design domains.

\begin{acks}
This work was supported by the Technology Innovation Program (RS-2025-02317326, Development of AI-Driven Design Generation Technology Based on Designer Intent) funded by the Ministry of Trade, Industry \& Energy (MOTIE, Korea).
\end{acks}

\appendix

\section{Details of Sequential Design Action Classification Framework}

\subsection{Data Collection and Preprocessing}
We analyzed designers' processes by abstracting raw interaction logs into high-level \textit{Design Actions} using a rule-based Event Log Abstraction (ELA) approach\cite{van2021event, mannhardt2016low, Baier_2014, van_Cruchten_2018}. Adopting an artifact-centric perspective\cite{van2021event, mannhardt2016low}, we tracked each artifact's lifecycle from creation through elaboration, relocation, and deletion to connect low-level events to higher-level design strategies. The abstraction removed operational noise and merged micro-events into discrete actions using time-based aggregation rules and artifact state changes. Our pipeline processed a total of 14,075 low-level events. 

In Step 1 (Event Cleaning), 6,782 events (48.2\%) were excluded, yielding 7,293 clean events. Exclusions comprised out-of-scope protocol violations and interview-date-related records (588), system noise such as auto-saves and internal updates (3,495), redundant duplicates and no-change logs (2,076), typing consolidation (460), and post-delete reappearing artifacts (163). 

In Step 2 (Action Extraction), we applied the design action taxonomy to the remaining 7,293 events and filtered out 1,455 events without design significance, resulting in 5,838 analyzable \textit{Design Actions}\cite{Baier_2014, van_Cruchten_2018}. This refinement reflects corrections to previously misclassified categories (e.g., \textit{Elaborate, ImageEdit, and IntentEdit}), leading to a stricter definition of meaningful design activity. In total, 8,237 events (58.5\%) were excluded across both stages.

\subsection{Action Magnitude Quantification Thresholds} 
Building on the deltas extracted in Stage~1 (Change Detection; Section~3), we quantified action magnitude using pilot-derived thresholds. Each continuous-valued action was classified into four magnitude levels based on data-driven quartile boundaries ($Q1$, $Q2$, $Q3$) of the pooled empirical distribution.
For \textit{Relocate} actions ($N = 3{,}168$), Euclidean displacement was partitioned at $Q1 = 87$~px, $Q2 = 302$~px, and $Q3 = 871$~px, yielding four levels: Micro ($\leq 87$~px; fine positional adjustments), Small ($88$--$302$~px), Medium ($303$--$871$~px), and Large ($> 871$~px; substantial reorganization). For \textit{Elaborate} text actions ($N = 196$), character-count change ($|\Delta_{\text{char}}|$) was partitioned at $Q1 = 3$, $Q2 = 12$, and $Q3 = 50$~characters, yielding four levels: Micro ($\leq 3$~chars; typo-level edits), Small ($4$--$12$~chars), Medium ($13$--$50$~chars), and Large ($> 50$~chars; paragraph-level revisions).

\subsection{Active Time and Segmentation}
Because sessions spanned multiple days, we standardized \textit{active time} by removing interaction gaps longer than 30~min. This cutoff is supported by the interval distribution ($95\%$ 3.93~min; $99\%$ 368.06~min) and sensitivity analyses, retaining short breaks while excluding long interruptions (e.g., meals/sleep) consistent with web-analytics sessionization. Active time was $3.73 \pm 1.75$~h (Baseline) and $4.03 \pm 2.13$~h (Agent$_{\text{organizer}}$), with no condition difference (Wilcoxon signed-rank $W = 16.0$, $p = 0.8438$).
We split each session into tertiles of cumulative active time—Early ($0$--$33.33\%$), Mid ($33.34$--$66.66\%$), Late ($66.67$--$100\%$). This equal-interval segmentation serves two purposes: First, it normalizes individual pace differences across participants with variable session durations, enabling cross-participant comparison without imposing fixed-duration phase boundaries. Second, it captures the temporal evolution of design strategies, reflecting the theoretically established progression from divergent exploration through structuring to convergent refinement in design processes~\cite{cross2023design, dorst2001creativity}. The observed phase-specific behavioral patterns (Section~4.3) provide post-hoc evidence that the tertile boundaries captured meaningful transitions in design strategy.

\subsection{Concurrent Event Handling}
A single user interaction can produce multiple events at the same time—for example, when an artifact is moved and its content is modified in one operation. Because these co-occurring events are recorded in an arbitrary order, naively treating that order as ground truth would introduce directional bias into transition probability estimates. To address this, we treated all orderings of a concurrent pair as equally likely. Specifically, given a concurrent pair [A, B] flanked by a preceding action X and a following action Y, we expanded the sequence into two permutations—X $\rightarrow$ A $\rightarrow$ B $\rightarrow$ Y and X $\rightarrow$ B $\rightarrow$ A $\rightarrow$—each weighted at 0.5. When an n-gram window contained multiple concurrent pairs, we enumerated all possible permutations (2 raised to the number of concurrent pairs) and assigned each an equal weight so that their combined contribution summed to one. This permutation-based weighting was applied to n-gram pattern extraction and to the Markov transition matrices used in edge-level analysis; frequency-based analyses such as action counts relied on the original raw counts without modification. 

\section{Analysis Strategy}
We conducted three analyses to address RQ1--RQ3 and examine workflow transformations across conditions.

First, \textbf{Design Workflow Redistribution Without Increased Effort (RQ1)} focused on understanding the overall behavioral reconfiguration following AI organizing agent introduction and validating efficiency. We analyzed distributional differences in action categories across conditions using Pearson's Chi-square test ($\chi^2(\textit{df} = 10, N = 5{,}838)$, $\alpha = 0.05$). Post-hoc analysis calculated standardized residuals for each action category to derive Z-scores. To confirm that observed distributional changes did not merely stem from task-time variations, we conducted a parallel total active time comparison by the Wilcoxon signed-rank test to verify experimental equivalence ($n = 8$ paired samples, two-tailed). Additionally, sensitivity analysis excluding \textit{AgentGen} behaviors confirmed that the AI organizing agent's impact represented an overall strategic shift rather than simply adding new behavior types.

Second, \textbf{Sequential Design Action Mining (RQ2)} was performed to identify design behavior patterns emerging during AI organizing agent collaboration. We extracted consecutive 2--3 action sequences (bigrams and trigrams) from the complete behavioral sequence across the entire session and calculated their relative frequencies within each condition. To statistically compare pattern frequencies between conditions, we applied two-proportion Z-tests to each bigram and trigram, testing whether the observed proportion of a given pattern differed significantly between Baseline and Agent$_{\text{organizer}}$. This analysis primarily focused on discovering discriminative patterns across conditions. We defined condition-specific patterns as sequences absent from Baseline's top 10 patterns but present in Agent$_{\text{organizer}}$'s top 10 patterns. We particularly focused on AI-mediated loops and self-repetition patterns involving \textit{AgentGen} and \textit{Create} to analyze 'generate-and-curate' collaboration dynamics. 

Third, \textbf{Modeling Phase-dependent Externalization (RQ3)} was designed to identify the evolution of the AI organizing agent's role over time. We divided workflows into tertiles of cumulative active time—Early ($0$--$33.33\%$), Mid ($33.34$--$66.66\%$), Late ($66.67$--$100\%$)—to normalize pace differences and capture temporal shifts in design strategy~\cite{cross2023design, dorst2001creativity}. Within each phase, we modeled designers' AI-mediated externalization through three complementary analyses. (1) We computed transition probabilities using first-order Markov chains~\cite{McComb_2018}, representing behavioral transitions as lag-1 sequential pairs (action$_i$ $\rightarrow$ action$_j$), and calculating conditional probabilities as follows: $P(\text{action}_j \mid \text{action}_i) = \frac{\sum w(\text{action}_i \rightarrow \text{action}_j)}{\sum_k \sum w(\text{action}_i \rightarrow \text{action}_k)}$, where $w$ denotes the transition weights that account for concurrent events (see Section~A.4 for details). Transition probabilities were calculated preserving designer boundaries (i.e., excluding cross-designer transitions). (2) We computed phase-level agent-originated canvas addition rates to measure how selectively designers integrated agent-generated outputs. This rate represents the proportion of artifacts placed on the canvas that originated from the agent (Create\_inspiration\_agent) among all artifacts added immediately following \textit{AgentGen} interactions, complementing the Markov transition probability $P(\textit{Create} \mid \textit{AgentGen})$. (3) We tracked action-rank shifts by ordering design action categories by their relative frequency within each phase and comparing rank positions across conditions, thereby identifying changes in the relative prominence of each action type as the design process unfolded.

\bibliographystyle{ACM-Reference-Format}
\bibliography{reference}
\end{document}